%% file: stefano25_v5.tex
\documentclass{elsart}
\usepackage{epsfig}
\usepackage{procrpc4}
\begin{document}
%%%%%%%%%%%%%%%%%
%%% MY Definitions
\def \NN{$N\overline{N}~~$}
\def \DD{$D^0\overline{D^0}~~$}
\def \D0bar{$\overline{D^0}~~$}
\def \ee{$e^{+}e^{-}~~$}
\def \blankline{\par\vskip 12 pt\noindent}
%%%%%%%%%%%%%%%%%%
\begin{frontmatter}
 \begin{flushright}
   FERMILAB-Pub-01/081-E \\
   FRASCATI-LNF/01-022(P)
 \end{flushright}
 \title{ 
   Evidence for a narrow dip structure at 1.9~GeV/c$^2$ 
   in $3\pi^+ 3\pi^- $ diffractive photoproduction
 }
%
% author list
%
 \input{author_list_plb.tex}

%
% Abstract
%
 \begin{abstract}
  A narrow dip structure has been observed at 1.9~GeV/c$^2$
  in a study of diffractive photoproduction of
  the $~3\pi^+3\pi^-$ final state performed by the Fermilab experiment E687. 
  \\
%  \begin{center}
%   DRAFT 7.21  Jun 8 2001 includes Rosner's references. and claes now at
%  \end{center}
 \end{abstract}
\end{frontmatter}

\section{Introduction.}
\vskip 3 truemm

The Fermilab Experiment 687 collaboration has collected
a large sample
of high-energy photoproduction
events, recorded with the E687 spectrometer
\cite{Frabetti:1992au}\cite{Frabetti:1993bn}
during the 1990/91
fixed-target runs at the Wideband Photon beamline at Fermilab. 
Although the experiment is focussed on charm physics,
a very large sample of diffractively photoproduced light-meson events 
was also recorded. 
This paper reports on a study 
of the diffractive photoproduction of the $3\pi^{+}3\pi^{-}$ final state
and the 
observation of a narrow dip in the mass spectrum at 1.9~GeV/c$^2$.
\par
\section {E687 spectrometer}
%
%ck I GOT RID OF A PIECE HERE -REPETITIVE
%
In E687, a forward multiparticle spectrometer is used to
measure the interactions of high-energy photons on a 4-cm-thick Be target.
It is a large-aperture, fixed-target spectrometer with
excellent vertexing, particle identification, and reconstruction
capabilities for photons and $\pi^0$'s. The photon beam is
derived from the Bremsstrahlung of secondary electrons 
of  $\approx 300$ GeV endpoint energy, which were produced by the 800
GeV/$c$ Tevatron proton beam. The charged particles emerging from
the target are tracked by a system of twelve planes of silicon microstrip
detectors arranged in three views.~These provide high-resolution separation
of primary (production) and secondary (charm decay or interaction) 
vertices. 
%The system lies downstream of the target and consists of twelve
%planes of microstrips arranged in three views. 
The momentum of a charged particle
is determined by measuring its deflections in two analysis magnets of
opposite polarity with five stations of multiwire proportional
chambers. Three multicell threshold \v Cerenkov counters are used to
discriminate between pions, kaons, and protons. Photons and neutral pions
are reconstructed by electromagnetic (EM) calorimetry. Hadron calorimetry 
and muon detectors
provide triggering and additional particle identification.
\par
\section{Event Selection}
\begin{figure}
\centerline{\protect
\epsfig{file=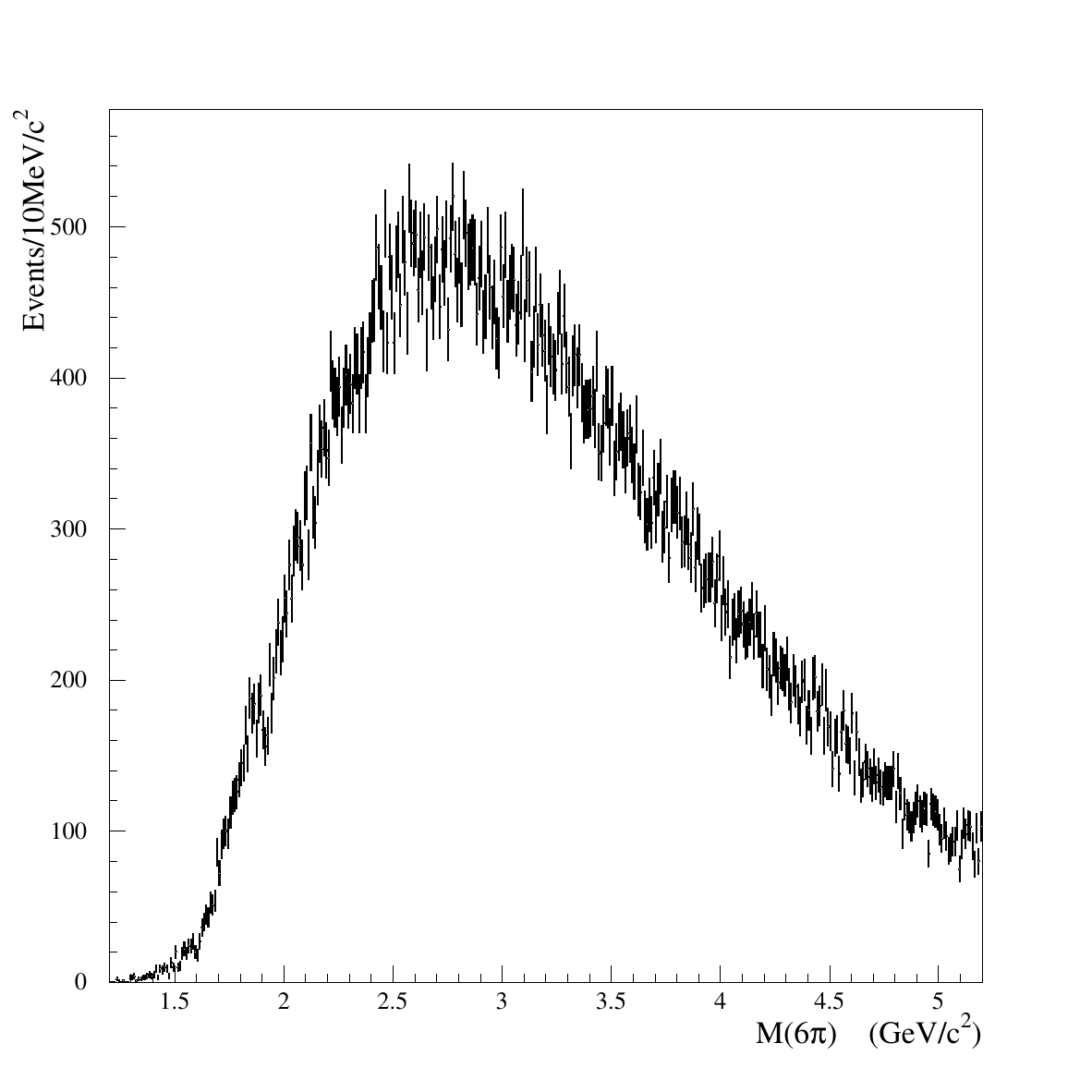,width=4.0in,height=3.0in}
}
\fcaption{ Distribution of $3\pi^+ 3\pi^- $ invariant mass after applying
a cut on the total\\ energy deposited in the calorimeters with
respect to the total energy in the \\ spectrometer.} 
\label{fig:first}
\end{figure}
\vspace{0.1in}
\par
Pions are produced in photon interactions in the Be target.
%.
The data acquisition trigger requires a minimum energy deposition in the hadron
calorimeters %(HC and CHC) 
located behind the electromagnetic calorimeters and
at least three charged tracks outside the pair region. 
The microstrip system and the forward spectrometer 
measure the 6$\pi$  final state (in this paper, 6$\pi$ refers to the
$3\pi^+ 3\pi^- $ state) 
with a mass resolution $\sigma =  10 ~{\rm MeV/c^2}$ at a total 
invariant mass of about 2~GeV/c$^2$.
It is required that a single six-prong vertex be reconstructed in the
target region by the microstrip detector, with a good confidence level.
Such a requirement rejects  background due to secondary interactions 
in the target. 
Exclusive final states are selected by also requiring that the same
number of tracks be reconstructed in the magnetic spectrometer.
The six microstrip tracks and the six spectrometer tracks are required 
to be linked, with no ambiguity in associating the microstrip and 
spectrometer tracks.
Events with particles identified by the {\v C}erenkov system 
as definite electrons, kaons, or protons, or as kaon/proton ambiguous are
eliminated and at least four out of 
six particles have to be positively identified as $\pi^{\pm}$.
Particle identification is tested by assuming that one or two out of the 
six tracks is a $K^{\pm}$, by computing all two-track invariant mass
combinations, and verifying  that there is no evidence of a peak
at the $K^{*}$ or at the $\phi$ mass.
We eliminated final states with $\pi^0$'s by rejecting events with
visible energy in the electromagnetic calorimeters that was not associated
with the charged tracks. A cut in this variable
($E_{\rm cal}/E_{6\pi} ~\leq~5\%$) is applied 
on the calorimetric neutral energy normalized to the
six-pion
energy measured in the spectrometer. 
The distribution of the 
six-pion invariant mass after these 
cuts is shown in Fig.~1. The plot shows a structure 
at 1.9~GeV/c$^2$. In the following, 
only the 6$\pi$ mass region around this structure will be analyzed.
\par
For diffractive reactions at our energies, 
the square of the four-momentum transfer $t$ 
can be approximated by the square of the total transverse momentum $P_{T}^2$
of the diffractively produced hadronic final state. 
Using this definition, the $P_{T}^2$ distribution of
diffractive events, Fig. 2, is well described by two
exponentials:  a coherent contribution with a slope
b$_c = 54\pm 2~$(GeV/c)$^{-2}$
consistent with the Be form factor 
 \cite{DeJager:1987qc}
and  an incoherent contribution 
with a slope b$_i = 5.10\pm 0.25 ~$(GeV/c)$^{-2}$.
\begin{figure}
\centerline{\protect
\epsfig{file=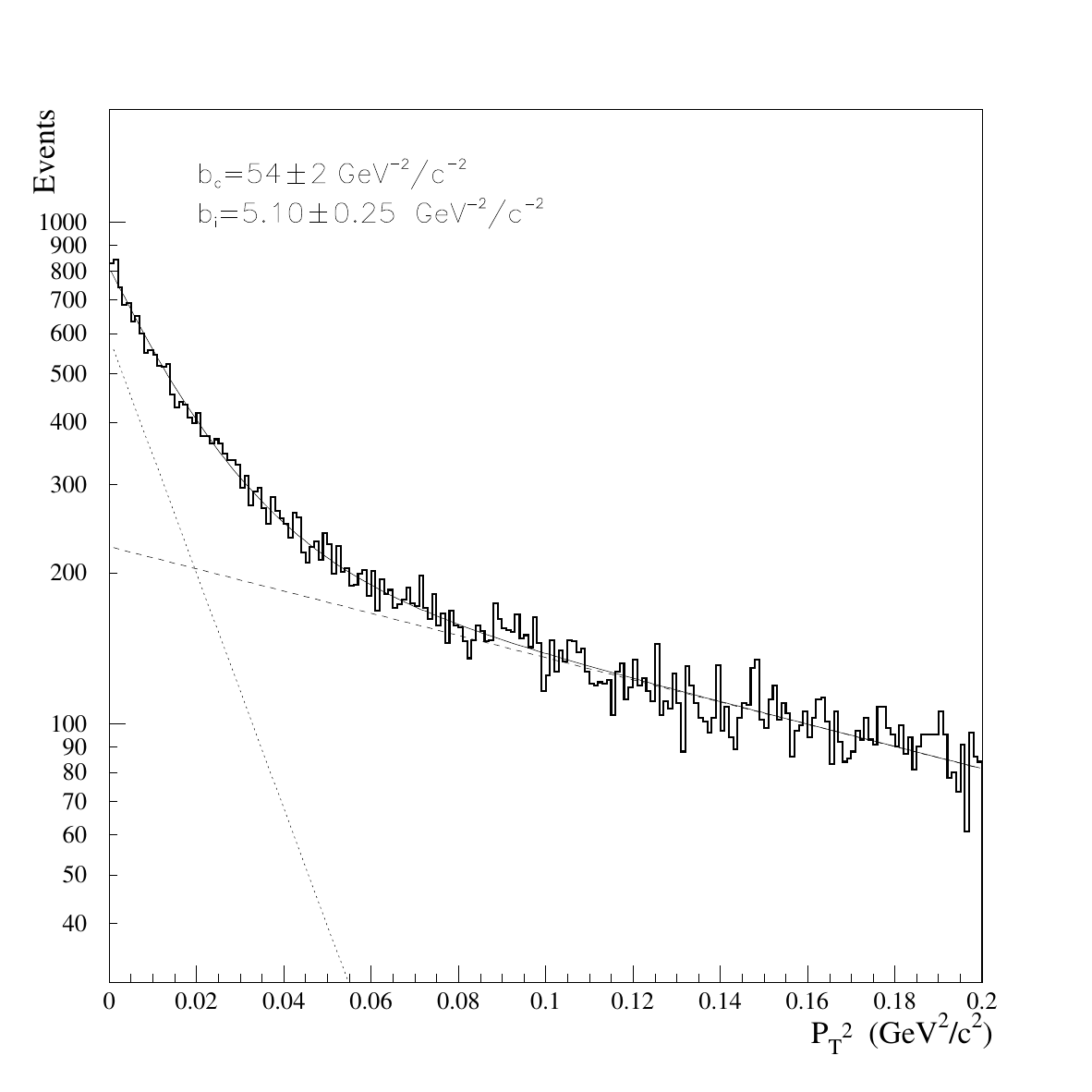,width=4.0in,height=3.0in} }
\fcaption{Tranverse momentum squared distribution showing the diffractive
peak. 
% Dotted distributions show coherent and incoherent contributions.
} 
\label{fig:second}
\end{figure}
Taking only events with $P_{T}^2 \leq 0.040~ $GeV$^2/c^2$, we evaluated
a contamination of about $ 50 \%$ from nondiffractive events. 
This incoherent contribution shows no structure in the
1.2$-$3.0~ GeV/c$^2$ mass range, Fig. 3.
The diffractive mass distribution was obtained by subtracting this
contribution, parametrized by a 
polynomial fit, and dividing the yield by the detection efficiency.
\par
The detection efficiency was computed by
modeling diffractive photoproduction of a mass M, using the 
experimentally found
slope b$_c$, and simulating the decay M$ \rightarrow{6\pi}$ 
according to phase space \cite{genbod}.
There is no threshold or discontinuity for the efficiency, particularly in the 
region of the dip structure. At $1.9 ~{\rm GeV/c}^2$, the (self-normalized
relative) efficiency {\it A} varied as 
{\it dA/A/dM}$_{6\pi}$ = 0.15/$~{\rm GeV/c}^2$.
The efficiency and the efficiency-corrected distribution of 
the six-pion invariant mass for
diffractive events, 
in the mass range 1.4$- 2.4~{\rm GeV/c}^2$, are shown in Fig.~4.
There  no evidence, albeit with large combinatorial backgrounds, for
resonance substructure, e.g., $\rho^0 \rightarrow \pi^+ \pi^-$, in the 6$\pi$
data below $M_{6\pi} = 2.0~{\rm GeV/c}^2$, either at the mass region of the dip
or in nearby sidebands. Similarly, the efficiency or
acceptance exhibited no threshold, edge, or discontinuity
over the entire mass region observed, when the six pion state
was simulated as a sequence of decays of intermediate two-body
resonances, for example $a_1^+ + a_1^- \rightarrow (\rho^0 \pi^+) + (\rho^0 \pi^-)
\rightarrow (\pi^+ \pi^- \pi^+) + (\pi^+ \pi^- \pi^-)$, even under extreme
assumptions  of full longitudinal or transverse polarizations 
for the initial state.
\par
\begin{figure}
\centerline{\protect
\epsfig{file=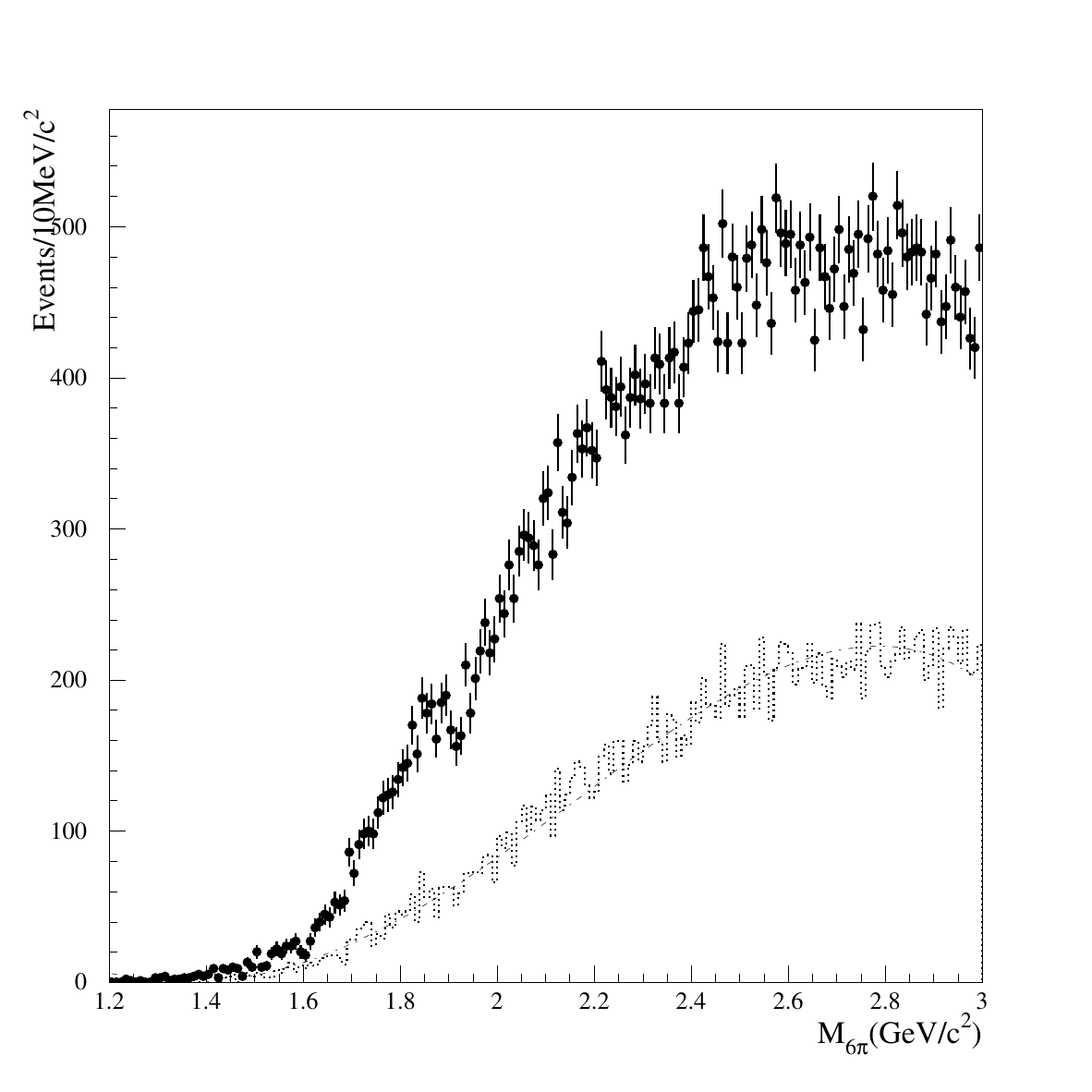,width=4.0in,height=3.0in}
}
\fcaption{ Distribution of $3\pi^+ 3\pi^- $ invariant mass in the 
1.2$-$3.0~ GeV/c$^2$ mass
range:\\ coherent plus incoherent contribution. 
Dotted distribution: incoherent\\ contribution.}
\label{fig:third}
\end{figure}

The presence of a dip at 1.9~GeV/c$^2$ was verified by
several checks of the systematics.
Diffractive photoproduction of \DD 
pairs or the associated production of  
\D0bar plus a charm baryon at low t 
(where the decay products of the other 
charm meson or charm baryon are missed) followed by the 6 $\pi$ decay
of the D$^0$ or \D0bar
are estimated  by Monte Carlo simulation to be negligible contributions.
It was also checked that demanding more stringent  cuts 
(i.e., requiring that all six particles be identified
as $\pi^{\pm}$, applying a sharper cut 
on the calorimeter neutral energy, or subtracting the
incoherent contribution bin by bin) increases the statistical
errors without significantly affecting the behavior shown in
Fig.~\ref{fig:third}.
\par
A three-parameter polynomial 
fit was performed, solid line in Fig.4, 
to explore the hypothesis that any structure in this distribution 
is a statistical fluctuation.
The normalized residual distribution,
evaluated for each 10-MeV/c$^2$ bin, 
is good in the full invariant mass range (Fig.5: left),
with the exception of the 
interval centered at 1.9~GeV/c$^2$, the region
of the claimed dip, where a poor $\sim~ 10^{-3}$ confidence level 
interval  (Fig. 5: right), is obtained,
making it highly unlikely that the observed dip is a statistical fluctuation.
Incoherently adding a Breit-Wigner to the fit does not improve the fit
quality much, dashed lines in Fig.~4 and Fig.~5.
\par
\section {Fitting the six-pion invariant mass distribution} 
\begin{figure}
\centerline{\protect
\epsfig{file=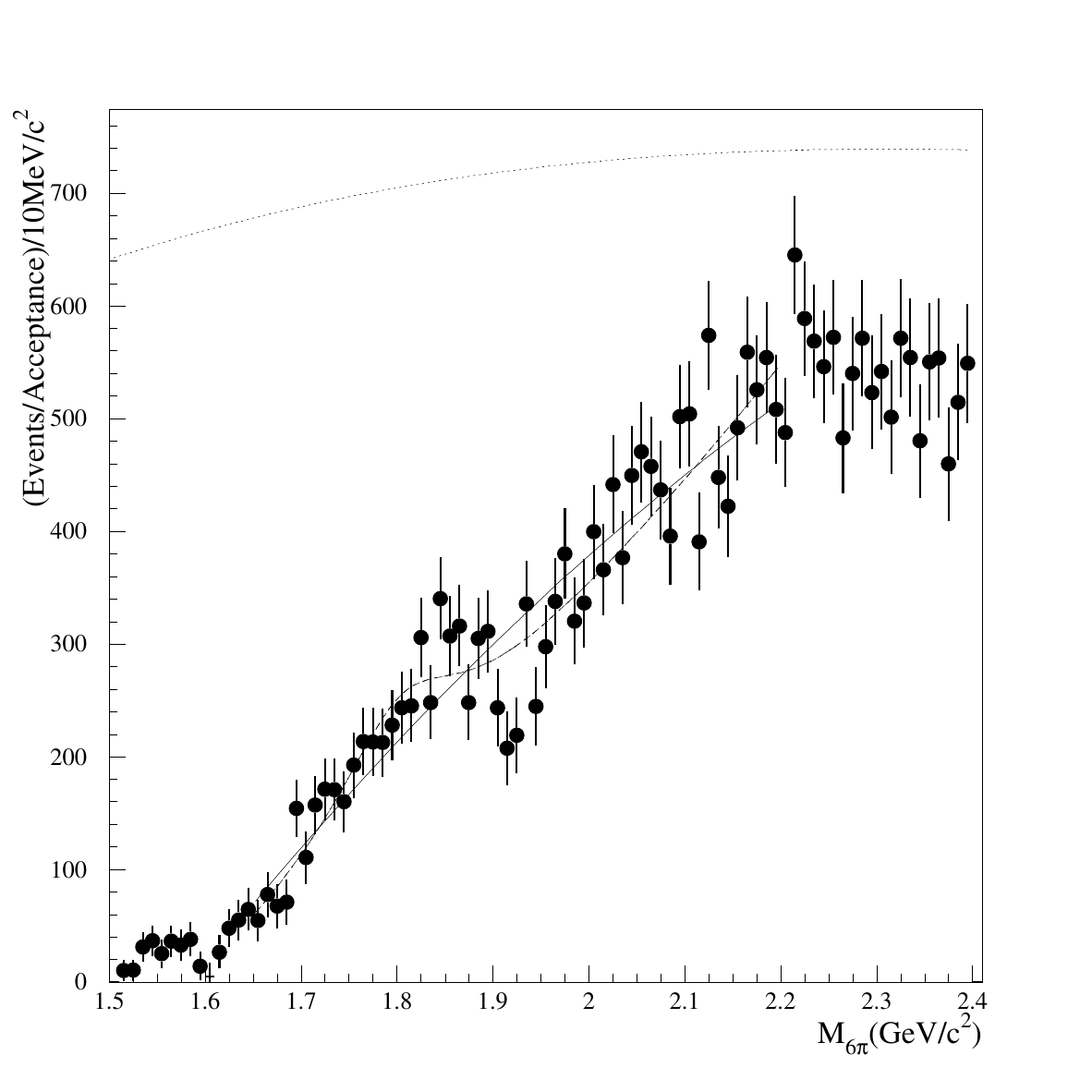,width=4.0in,height=3.0in}
}
\fcaption{Acceptance-corrected distribution of $3\pi^+ 3\pi^- $ invariant mass
for diffractive\\ events after subtracting incoherent contribution.
Solid line: second-order \\polynomial fit. Dashed line: polynomial fit
with incoherently-added Breit-Wigner.\\
Upper dot line: relative detection efficiency (arbitrary normalization).}

\label{fig:fourth}
\end{figure}

\par\noindent

\begin{figure}
\centerline{\protect
\epsfig{file=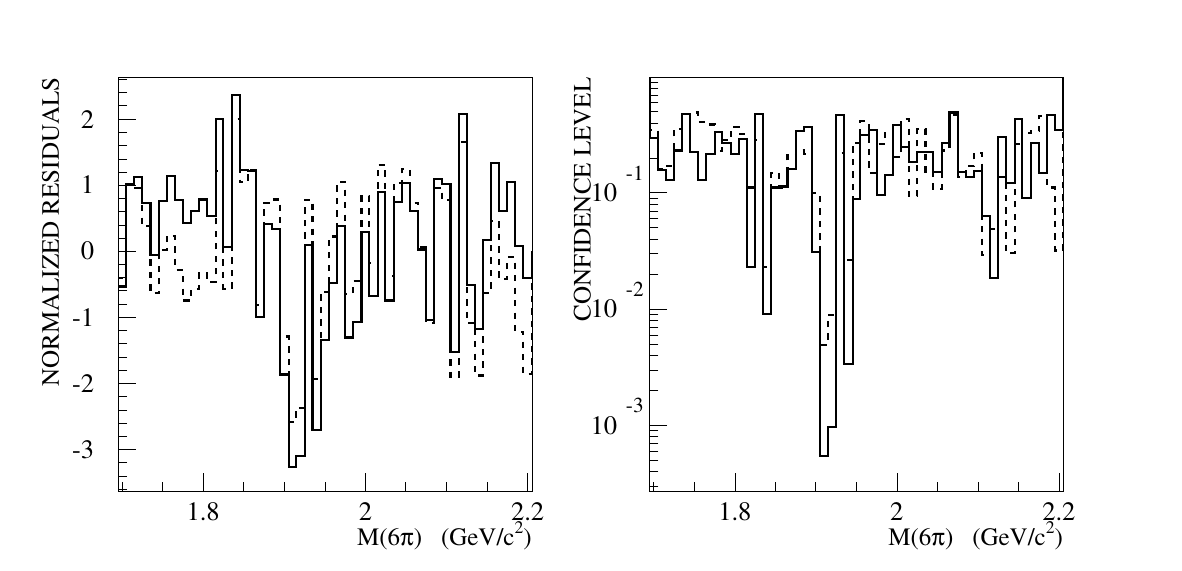,width=5.0in}
}
%\fcaption{ Confidence level for any 50 MeV/c$^2$ mass range intervals. 
%Solid line : second-order polynomial fit.
%Dashed line : second
%order polynomial with incoherently Breit-Wigner fit.}
\fcaption{ Normalized residual (left) and confidence level (right)
  distributions.
\par\noindent
Solid line: second-order polynomial fit.Dashed line: Breit-Wigner plus 
second-order \\ polynomial fit.}
\label{fig:fifth}
\end{figure}
Because of the narrow width, the E687 spectrometer mass
resolution, $\sigma = 10 ~{\rm MeV/c}^2$ at 2 GeV/c$^2$,
was unfolded 
by applying the method described in Ref.~\cite{NIM}:
the experimentally observed data distribution $r(x)$ and the unfolded mass
distribution $a(x)$ are related by
$a(x)~\sim~r(x)~-~0.5 \sigma^2 \cdot r(x)^{''}$, where r(x)$^{''}$ is the
second derivative with respect to M$_{6\pi}$ for the observed distribution.
This relationship results from applying a Fourier transform
and approximating the resolution function by 
$g(x)~=~ exp(-\frac{\sqrt{2}|x|}{\sigma})$, which is
marginally different from a Gaussian.
A fit similar to the following one is used to obtain $r(x)^{''}$ from the unfolded data. 
 The data after unfolding are
shown in Fig. 6.
\par
The dip structure at 1.9~GeV/c$^2$ has been characterized by a two-component
fit, adding coherently a relativistic Breit-Wigner resonance to a
diffractive continuum contribution.
The continuum probability distribution \(F_{JS}(M)\) has been modeled after 
a Jacob-Slansky diffractive parameterization\cite{jacob}, plus a constant term
$c_0$

 \[ F_{JS}(M) = f_{JS}^2(M) = c_0 + c_1 \frac{e^{\frac{-\beta}{M-M_0}}}{(M-M_0)^{2-\alpha}}
  \]

The Jacob-Slansky model is based on probabilities, rather than amplitude
phenomenology, and specifically has no prescription as to how the complex
phase of the continuum distribution varies with  mass. 
The Jacob-Slansky amplitude \(f_{JS}(M)\) is assumed to be the purely
 real  ($\phi_{JS} \equiv 0$) square root of the probability function 
 \( F_{JS}(M) \).  For the fit, a
relative phase factor $e^{i\phi}$, independent of mass, and a normalizing 
factor $a_r$ multiplied a relativistic Breit-Wigner resonance term, giving 
the overall amplitude

\[ A(M) = f_{JS}(M) + a_r \frac{-M_r \Gamma e^{i\phi}}{M^2-M_r^2 +
 i M_r\Gamma} \]

\begin{figure}[ht]
\centerline{\protect
\epsfig{file=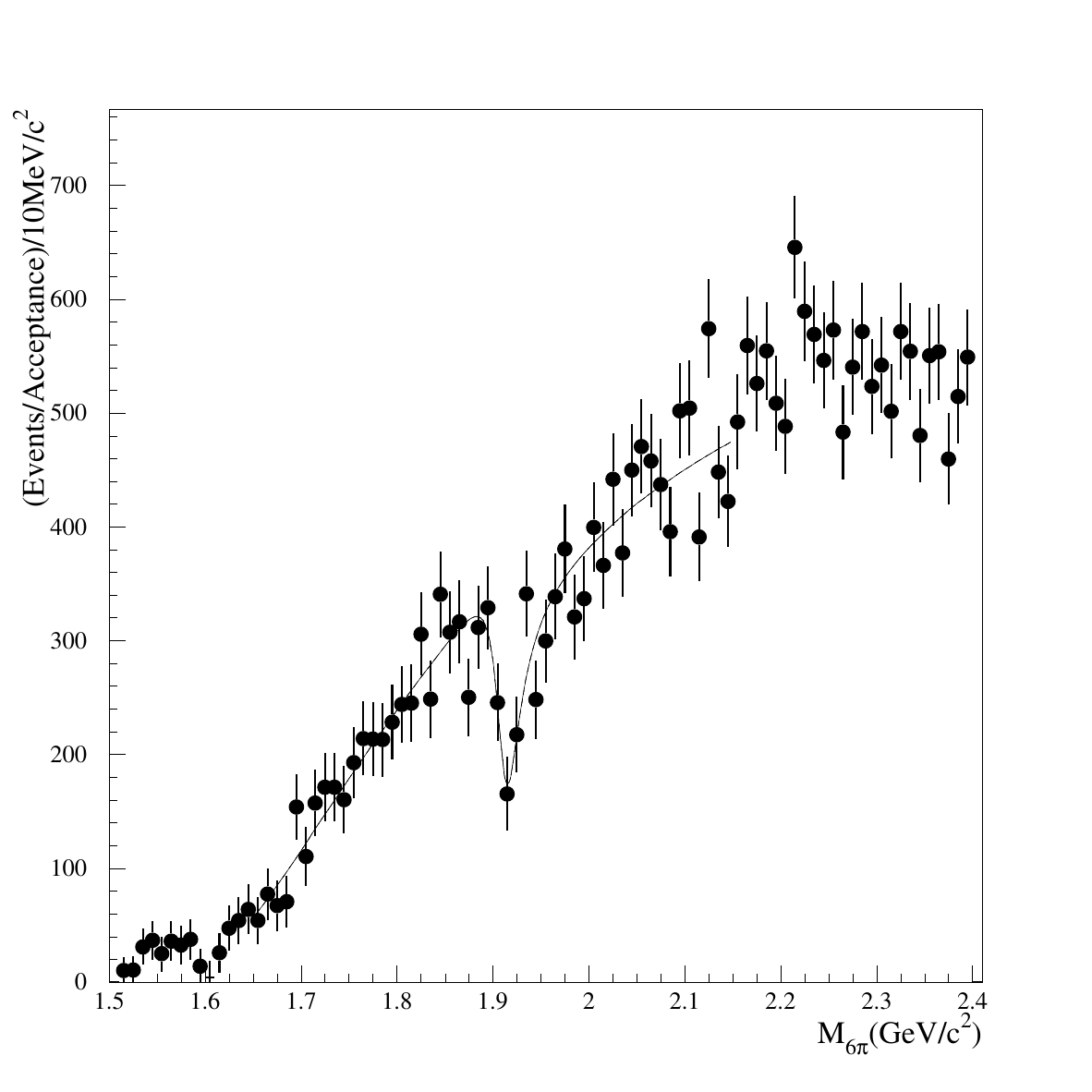,width=4.0in,height=3.0in}
}
\fcaption{ Acceptance-corrected distribution of $3\pi^+ 3\pi^- $ invariant
mass for diffractive \\ events. 
The mass resolution has been unfolded. Fit parameters are listed in \\
Table 1.} 
\label{fig:sixth}
\end{figure}

Fit results are shown in Table~\ref{tab:unfolded} and in
Fig.~\ref{fig:sixth} for a fitted mass range
from 1.65 to 2.15 {\rm GeV/c}$^2$,
symmetric with respect to the dip. Quantities shown are
the mass and width of the resonance, the amplitude ratio
 $a_{\rm r}/f_{\rm{JS}}(M_{\rm r})$  between the Breit-Wigner function and
 the Jacob-Slansky continuum, the 
relative phase and the $\chi^2$/dof. 
The Jacob-Slansky parameters for that one mass range are also given.
Fit values show consistent evidence for a narrow resonance at
M$_r$ ~=~1.911 $\pm$ 0.004 $\pm$ 0.001 ~GeV/c$^2$ with a width 
$\Gamma$ = 29 $\pm$ 11 $\pm$ 4 ~MeV/c$^2$,
where the errors quoted are statistic and systematic,
respectively. 
The fit values shown in Fig.~6 and represented by the parameters
of Table 1 are stable with acceptable $\chi^2/{\rm dof}$ over varying mass
ranges from 1.65 to 2.3 {\rm GeV/c}$^2$.
We quote as systematic error the sample variance of the fit values 
due to our choice of fit mass range. 
The  quality of the fit deteriorates somewhat as the upper limit
of the fit for this simple model is extended from
2.1 to $3.0~{\rm GeV/c^2}$.
%as can be expected given the simplicity of the model. 
However, the only fit parameter that is affected is the
width, which varies from $29\pm 11 \, {\rm MeV/c}^2$ to 
$40 \pm 20 \, {\rm MeV/c}^2$.
\par
\begin{table}
\caption{Fit results for a mass range from 1.65 to 2.15 $GeV/c^2$
\label{tab:unfolded}
}
\begin{center}
\begin{tabular} {|l|r|l|r|} 
\hline\hline
M$_r$ (GeV/c$^2$) & $1.911 \pm 0.004$ \\
$\Gamma$ (MeV/c$^2$) & $29 \pm 11$ \\
%A$_{\rm r}$/A$_{\rm cont}$ & $0.31 \pm 0.07$ \\
$a_{\rm r}/f_{\rm{JS}}(M_{\rm r})$ & $0.31 \pm 0.07$ \\
$\phi$ (deg.) & $62 \pm 12$ \\
$\chi^2$/dof & 1.1 \\
$M_0$ (GeV/c$^2$) & $1.49 \pm 0.02$ \\
c$_0$ (GeV/c$^2$)$^{-1}$ & $0 \pm 50$ \\
c$_1$(GeV/c$^2$)$^{1-\alpha}$ & $960 \pm 80$ \\
$\beta$ (GeV/c$^2$) & $0.5 \pm 0.3$ \\
$\alpha$  & $1.8 \pm 0.2$ \\
\hline\hline 
\end{tabular}
\vfill
\end{center}
\end{table}
\par 
The dip structure is reminiscent of what was observed
% with  less
%statistical significance, 
in e$^+$e$^-$ annihilation \cite{baldini0}
\cite{Donnachie}.
The mechanism by which a narrow resonance may interfere destructively with
a continuum, or a broad resonance, could be similar to the one 
described in \cite{Gilman}, in a different context.
Vector ${\rm q} \overline{\rm q}$ hybrids are  predicted at 
$\sim$ 1.9 ~GeV/c$^2$
according to the flux tube model \cite{Isgur} \cite{Close}.
%and to recent lattice calculations \cite{UKQCD} \cite{Bernard} \cite{Lacock}.
 A hybrid is
expected to have a small, but not vanishing, e$^+$e$^-$ width and a total
width constrained by decay selection rules \cite{Swanson}.
Vector glueballs are expected at higher masses, according to the bag model
and to lattice calculations\cite{UKQCD1}\cite{Morningstar}. 
Narrow resonances have also been predicted near the \NN region, but have
never been 
conclusively found
\cite{Dov}\cite{Sha}\cite{Jaf}\cite{Rich}\cite{Myh}.
A \NN threshold effect might also produce a downward step in the amplitude, 
followed by a recovery\cite{Rosner:1974sn}\cite{Rosner:1968si}. 
These
resonances should also be
observed as a steep variation 
in the nucleon time-like form factor\cite{PS170} \cite{FENICE2}.

\par
\par
\section{Conclusions}
The diffractive photoproduction of  $3\pi^{+}3\pi^{-}$  
 has been studied by E687. Evidence has been found for a narrow structure 
near $M_{6\pi}$ = 1.9 GeV/$c^2$.  If this dip is characterized as the
destructive interference of a resonance with the continuum background, then
the parameters of this resonance would be 
M$_r$ ~=~1.911 $\pm$ 0.004 $\pm$ 0.001 ~GeV/c$^2$, 
with $\Gamma$ = 29 $\pm$ 11 $\pm$ 4 ~MeV/c$^2$.
Such a resonance could be 
assigned the photon quantum numbers (J$^{\rm PC}=1^{--}$) and G=+1, I=1 due
to the final state multiplicity. 
There is little 
understanding of the specific mechanism responsible for this destructive 
interference. In order 
to facilitate additional phenomenological studies, the data points of 
Fig.~6 are available \cite{data}. 
\par
%

% added by Peter H. Garbincius, November 1, 2018
\section{2018 Addendum:  Comparison of the 6 pion energy distribution with Montecarlo simulation}
In September 2018, 17 years after publication, a reader requested the energy distribuion of the $6\pi$ state,
which was not included in publication of 2001.  This distribuion was found in the Fermilab E687/E831 Internal Note
six\_pi1.ps, dated 9/15/2000.  The figure and descriptive text below as sent to the reader are appended here to 
provide documentation of this distribution.
\par
%$\vspace{0.5 cm}
%\epsfig{file=fig7_6pi_energy_spectrum.pdf,width=4.0in,height=5.0in}

\begin{figure}
\centerline{\protect
\epsfig{file=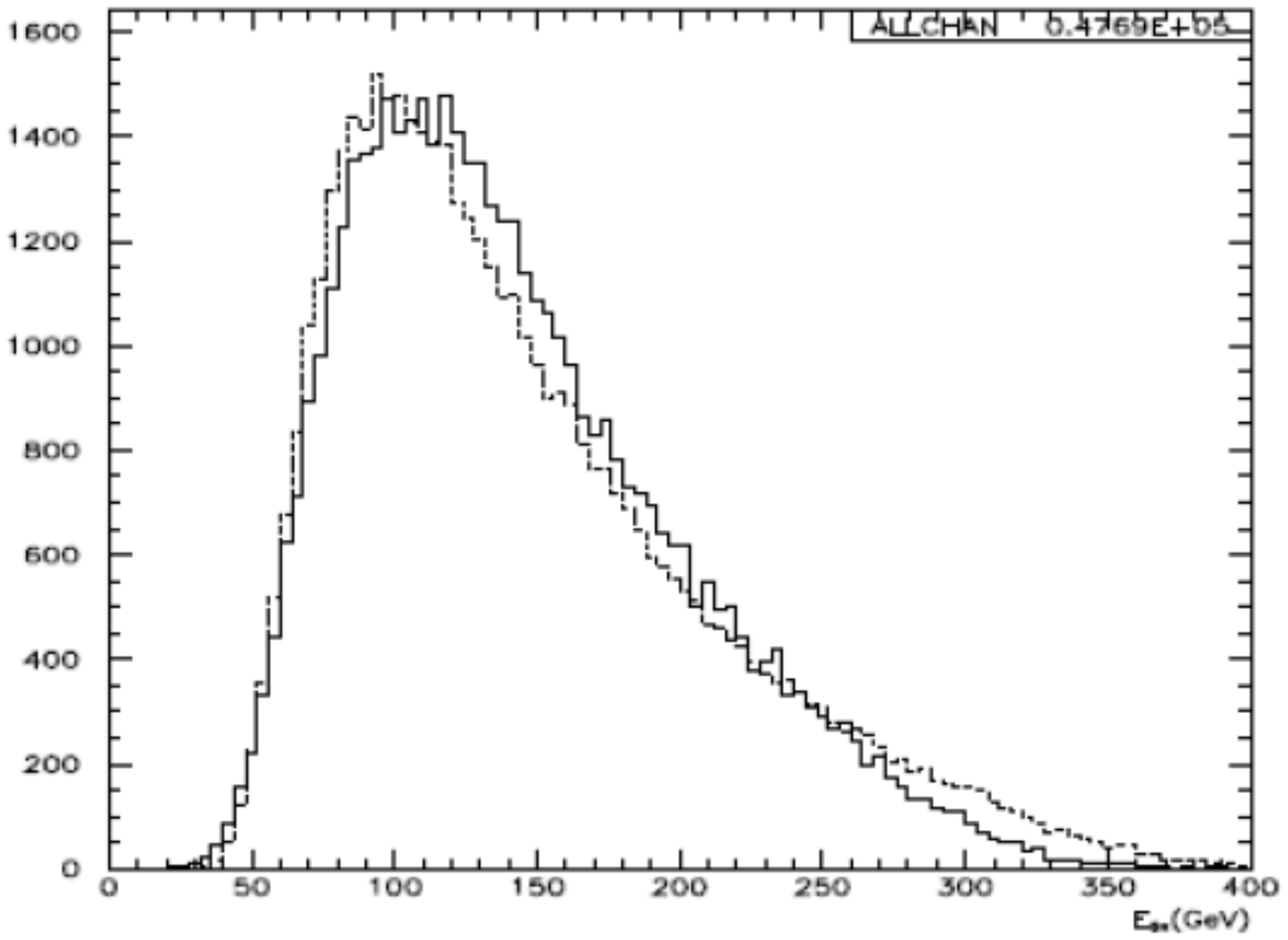,width=4.0in,height=3.0in}
}
\fcaption{ $3\pi^+ 3\pi^-$ energy distribution from E687 data (solid line) and E831 Montecarlo (dashed line).}
\label{fig:seventh}
\end{figure}
%$\vspace{0.5 cm}

To simulate E687 data, we have used the E831 Montecarlo with a top energy of 300 GeV.  Once generated,
the photon energy has been increased by a factor of 350/300 in order to take roughly into account the 
E687 beam energy.  So, our comparison of the energy distribution of our events with the E831 Montecarlo events
(Fig. 7) is only approximate.
\par

\vspace{1. cm}
We would like to thank the staffs of Fermi National
Accelerator Laboratory, INFN of Italy, and the physics departments
of the collaborating institutions for their assistance. 
This research was partly,supported 
by the U.~S. Department of
Energy, the U.~S. National Science Foundation, the Italian Istituto Nazionale di Fisica Nucleare and
Ministero dell'Universit\`a e della Ricerca Scientifica e Tecnologica,
the Brazilian Conselho Nacional de Desenvolvimento Cient\'{\i}fico e
Tecnol\'ogico, CONACyT-M\'exico, the Korean Ministry of Education, and
the Korean Science and Engineering Foundation. 
\par

\end{document}

%% file: author_list_plb.tex
\begin{center}

\normalsize

E687 Collaboration

\bigskip

P.L.~Frabetti$^a$,
H.W.K.~Cheung$^{b,1}$, 
J.P.~Cumalat$^b$, 
C.~Dallapiccola$^{b,2}$, 
J.F.~Ginkel$^{b}$               % Cumalat cant find, says use Colorado 6april, 
W.E.~Johns$^{b,3}$,             % Vanderbilt
M.S.~Nehring$^{b,4}$,           % Adams State College, Alamosa, CO 81102
E.W.~Vaandering$^{b,3}$,                % Vanderbilt
J.N.~Butler$^c$, 
S.~Cihangir$^c$
% ...
                                % South Korea
                        % Myung is still unknown - leave at Korea for now

\bigskip

\small

$^a$ {\it Dip.   di   Fisica   dell'Universit\`{a}   and   INFN-Bologna,  
I-40126   Bologna,   Italy.}

$^b$ {\it  University   of   Colorado,   Boulder,   CO   80309,  USA.}

$^c$ {\it  Fermi   National   Accelerator   Laboratory,   Batavia,  
  IL   60510,   USA.}

\end{center}
\begin{flushleft}
% ...
$^1$ Present address:  Fermi National Accelerator Laboratory, Batavia, IL
60510, USA. 

$^2$ Present address:  University of Massachusetts, Amherst, MA 01003, USA.

$^3$ Present address:  Vanderbilt University, Nashville, TN 37235, USA.

$^4$ Present address:  Adams State College, Alamosa, CO 81102, USA.

% ...
Science, Yusong P.O. Box 102, Taejon 305-600, South Korea. 

\end{flushleft}
\bigskip

$PACS:$ 13.25.Jx, 13.60.Le, 14.40.Cs

\bigskip